\documentclass[12pt,preprint]{aastex}

\slugcomment{Prepared for ApJ}
\shorttitle{}
\shortauthors{}

\begin{document}

\title{The Effect of Conjunctions on the Transit Timing Variations \\of Exoplanets}

\author{David Nesvorn\'y}
\affil{Department of Space Studies, Southwest Research Institute,
      1050 Walnut St., Suite 300, \\ Boulder, CO 80302,
      E-mail: davidn@boulder.swri.edu}
\author{David Vokrouhlick\'y}
\affil{Institute of Astronomy, Charles University,
       V Hole\v{s}ovi\v{c}k\'ach 2, CZ--18000 Prague 8, \\
       Czech Republic, E-mail: vokrouhl@cesnet.cz}

\begin{abstract}
We develop an analytic model for transit timing variations produced by orbital conjunctions between 
gravitationally interacting planets. If the planetary orbits have tight orbital spacing, which is 
a common case among the {\it Kepler} planets, the effect of a single conjunction can be best described 
as: (1) a step-like change of the transit timing ephemeris with subsequent transits of the inner 
planet being delayed and those of the outer planet being sped up, and (2) a discrete change in sampling 
of the underlying oscillations from eccentricity-related interaction terms. In the limit of small 
orbital eccentricities, our analytic model gives explicit equations for these effects as a function 
of the mass and orbital separation of planets. We point out that a detection of the conjunction effect in 
real data is of crucial importance for the physical characterization of planetary systems from
transit timing variations.     
\end{abstract}


\section{Introduction}

Planetary orbits are less than ideal clocks. This is because various processes, including, for example, 
the gravitational interaction between planets, collaborate to produce fluctuations from perfect periodicity. 
The Transit Timing Variations (TTVs; Agol et al. 2005, Holman \& Murray 2005), a method that has 
become increasingly important in the exoplanet research, exploits the effect of these fluctuations on
the timing of planetary transits. It can be used to make useful inferences about the nature of planets in a 
system where at least one planet is transiting.

The TTVs caused by interacting planets come in several flavors. The {\it long-periodic} TTVs result from 
orbital variability on timescales much longer than the orbital period. Their detection therefore requires 
a long observation baseline (Heyl \& Gladman 2007). The {\it resonant} and {\it near-resonant} TTVs occur 
when orbital periods, when divided by each other, are equal or nearly equal to a ratio of small integers. 
The planetary perturbations tend to build up in this situation, leading to TTVs with a large amplitude.

The detection of (near-)resonant TTVs is expected if a large set of transit 
observations is available, because some planetary systems are bound to have (near-)resonant 
orbits (either due to statistics or because they are driven to these orbits by formation processes). 
The description of planetary properties from (near-)resonant TTVs, however, is plagued with degeneracies, 
which may be resolved only under certain assumptions (Lithwick et al. 2012). This happens, in essence, 
because the (near-)resonant TTVs are limited in the information content.

This is where the {\it short-periodic} TTVs become useful. The short-periodic TTVs are produced by variations 
of orbits on a timescale comparable to the orbital period. In general, it can be shown that 
\begin{equation}
-n_i \delta t_i = \delta\lambda_i - 2\delta h_i + {3 \over 2}(k_i \delta h_i + h_i \delta k_i) + 
{\cal O}(e_i^2)  \ , 
\label{deltat}
\end{equation}
where $\delta t_i$ is the short-periodic deviation of timing of planet $i$ from a linear ephemeris, 
$k_i= e_i \cos \varpi_i$, $h_i=e_i \sin \varpi_i$, and $n_i=2\pi/P_i$, where $P_i$ is the orbital period 
(Nesvorn\'y 2009). Quantities $\delta \lambda_i$, $\delta h_i$ and $\delta k_i$ are the short-periodic 
variations of the mean longitude, $h_i$ and $k_i$, respectively.\footnote{The negative sign in front of Eq.
(\ref{deltat}) arises from the convention that a positive (negative) change of the mean longitude leads 
to negative (positive) $\delta t_i$. Also, it is assumed in Eq. (\ref{deltat}) that the observer's line 
of sight lies along the $X$ axis from which the orbital angles are measured.}

The short-periodic variations $\delta \lambda_i$, $\delta h_i$ and $\delta k_i$ can be computed from 
perturbation theory. With $m_i,a_i,e_i,i_i,\varpi_i, \Omega_i,\lambda_i$ denoting the mass and orbital 
elements of planet $i$, we have 
\begin{eqnarray}
\delta \lambda_i & = & {1 \over L_i} \left[ 2 a_i {\partial \chi \over \partial a_i} + 
{\beta \over D_i}\left( x_i {\partial \chi \over \partial h_i} -
y_i {\partial \chi \over \partial k_i} \right)\right]\ ,\nonumber \\
\delta h_i & = & -{\beta \over L_i} \left[ {\partial \chi \over \partial k_i} + 
{x_i \over D_i} {\partial \chi \over \partial \lambda_i} \right]\ ,\nonumber \\
\delta k_i & = & {\beta \over L_i} \left[ {\partial \chi \over \partial h_i} + 
{y_i \over D_i} {\partial \chi \over \partial \lambda_i} \right]\ ,
\label{lambda}
\end{eqnarray}
where $L_i = m_i \sqrt{G M_* a_i}$, $G$ is the gravitational constant, $M_*$ is the mass of the host star,
$x_i = -\sqrt{2 P_i} \sin \varpi_i$, $y_i = \sqrt{2 P}_i \cos \varpi_i$, $P_i = L_i(1-\beta_i)$, 
$D_i=\sqrt{2 L_i (1+\beta_i)}$ and $\beta_i = \sqrt{1-e_i^2}$. Function $\chi$ can be written as:
\begin{eqnarray}
\chi & = & {G m_1 m_2 \over a_2} \sum_{|k_1|+|k_2| \neq 0} \imath 
{C^{\mathbf{l},\mathbf{j}}_\mathbf{k}(\alpha) 
\over k_1 n_1 + k_2 n_2} e_1^{l_1} e_2^{l_2} (\sin {i_1 \over 2})^{j_1} 
(\sin {i_2 \over 2})^{j_2} \nonumber \\ &\times& 
\exp \imath (k_3 \varpi_1 + k_4 \varpi_2 + k_5 \Omega_1 + k_6 \Omega_2) 
\exp \imath (k_1 \lambda_1 + k_2 \lambda_2)
\label{chi}
\end{eqnarray} 
with $\imath=\sqrt{-1}$, $C^{\mathbf{l},\mathbf{j}}_\mathbf{k}(\alpha) = 
C^{\mathbf{l},\mathbf{j}}_{-\mathbf{k}}(\alpha)$, $\alpha = a_1/a_2 < 1$, 
and multi-indexes $\mathbf{l} = (l_1,l_2)$, $\mathbf{j} = (j_1,j_2)$ and
$\mathbf{k} = (k_1,k_2,k_3,k_4,k_5,k_6)$. See Nesvorn\'y \& Morbidelli (2008) for the assumptions
that led to the derivation of Eqs. (\ref{lambda}) and (\ref{chi}). In brief, Eq. (\ref{lambda})
does not include terms from the inclination of the transiting planet (assumed to be small), 
and Eq. (\ref{chi}) is given to the first-order in $m_1/M_*$ and $m_2/M_*$. 

According to these equations, $\delta t_i$ contains Fourier terms with the $k_1 \lambda_1 + k_2 \lambda_2$ 
harmonics. The amplitude of these terms is a complex function of $\alpha$, eccentricities and 
inclinations, but the ones with small $|k_1|$ and $|k_2|$ values are generally the most important 
(except if $k_1 n_1 + k_2 n_2 = 0$ for arbitrary $k_1$ and $k_2$, indicating 
the presence of a resonance). It is also obvious, in the approximation of Eqs. (\ref{deltat})-(\ref{chi}), 
that $\delta t_i$ is proportional to $P_i$, independent of $m_i$, and scales linearly with the 
companion mass.

The short-periodic TTVs are more difficult to detect observationally then the (near-)re\-sonant TTVs, 
because they generally have a small amplitude. If they are detected, however, they can be used to 
{\it uniquely} characterize the orbital properties of planets. This has been theoretically demonstrated 
in Nesvorn\'y \& Beaug\'e (2010) under the assumption that there is no a priori information 
about the (non-transiting) companion, and done in practice in Holman et al. (2010), and elsewhere. 
Intuitively, this can be understood because each Fourier term in Eq. (\ref{chi}) provides specific 
information about orbital elements. Thus, if at least a few of these terms are detected in real data, 
the information contained in the detection is high enough to make the inversion to orbital elements 
possible (e.g., Nesvorn\'y et al. 2012).  
 
\begin{figure*}
\epsscale{0.5}
\plotone{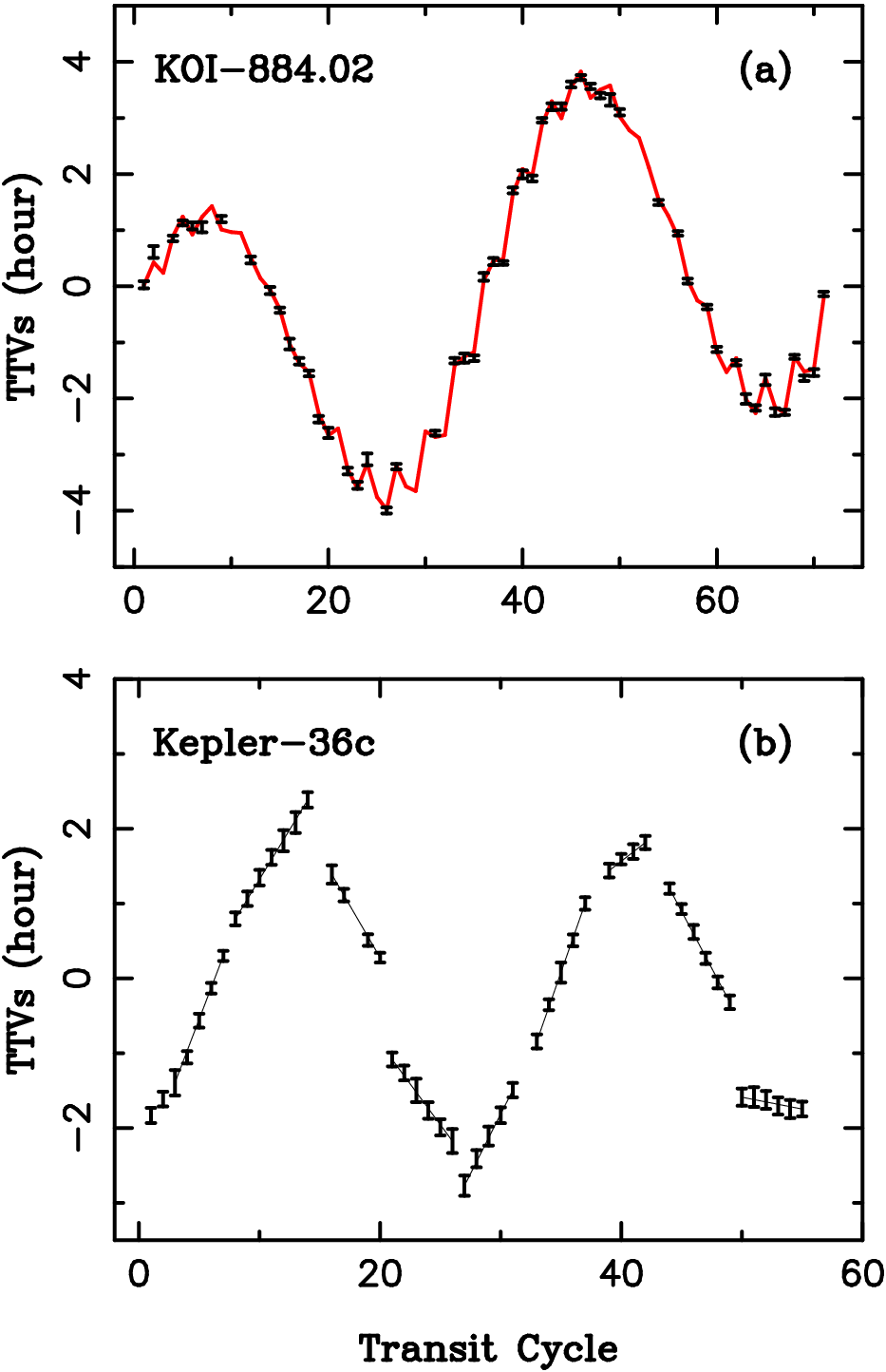}
\caption{The effect of conjunctions on the TTVs of: (a) KOI-884.02, and (b) Kepler-36c.
The TTV data for KOI-884.02 and Kepler-36c were obtained from Nesvorn\'y et al. (2014)
and Carter et al. (2012), respectively. The red line in panel (a) shows the best 
dynamical fit from Nesvorn\'y et al. (2014) corresponding to a (non-transiting) companion 
with the mass of $\simeq2.4$ $M_{\rm J}$ and outer orbit just wide the 3:1 orbital resonance with 
KOI-884.02. The TTVs of Kepler-36c, on the other hand, are caused by a transiting 
super-Earth (Kepler-36b) with an orbit in the 7:6 resonance with Kepler-36c. The line 
segments in panel (b) highlight the discontinuous nature of Kepler-36c's TTVs.}
\label{obs}
\end{figure*}

An important component of the short-periodic TTVs, which is the main focus here, is produced by 
{\it conjunctions} between planets. The conjunction effect can be conveniently illustrated using 
Kepler-36c (Carter et al. 2012) and KOI-884.02 (Nesvorn\'y et al. 2014) (see Figure \ref{obs}).
Kepler-36 is a doubly-transiting system, consisting of two planets on very tightly spaced orbits, 
with the two planets having significantly different masses (Carter et al. 2012). The KOI-884 system
contains three known transiting planetary candidates, with the inner two displaying little or 
no sign of TTVs, while the outer one (KOI-884.02) exhibits significant TTVs. Nesvorn\'y et al. (2014) 
used the TTVs of KOI-884.02 to detect an additional, unseen (i.e. non-transiting) planet just narrow
of the outer 3:1 resonance with KOI-884.02 (orbital period ratio 2.93).

In the case of KOI-884.02, the effect of conjunctions is best seen for transit cycles 32 to 45
in Figure \ref{obs}a, where three transits on a nearly linear ephemeris are offset from the next three 
transits. These discontinuities are produced at conjunctions with the outer massive planet 
($M_{\rm p}\simeq2.4\ M_{\rm J}$, where $M_{\rm J}$ is the mass of Jupiter; Nesvorn\'y et al. 2014). 
The TTVs of Kepler-36c, on the other hand, are a series of approximately linear segments that 
are tilted relative 
to each other at different angles (Carter et al. 2012). These discontinuities correspond to the 
orbital conjunctions between Kepler-36c and the inner transiting planet Kepler-36b. In this case, 
the very tight spacing of two orbits at (or very near to) the 7:6 resonance implies the physical 
distance between planets to be small near conjunctions, and large variations are therefore expected. 
In general, for a $(j+1)$:$j$ resonance, we expect the conjunctions to occur in every $j$ periods 
of the outer planet (and $j+1$ periods of the inner one). This is why, in Figure \ref{obs}b, groups 
of $j=6$ transits share the same linear ephemeris (note that some transits are missing due to 
instrumental and other issues).     

It is not straightforward to theoretically understand the effect of conjunctions from Eqs. 
(\ref{deltat})-(\ref{chi}), mainly because the TTV discontinuities occurring at conjunctions 
are difficult to approximate by the Fourier series, and because these equations include many different 
terms such that it is not clear which ones are responsible for the conjunction effect. The goal
of this paper is to present a simple model for the conjunction effect that can be used as an
intuitive guideline for more realistic modeling of cases such as the ones shown in Figure 
\ref{obs}. 

In Section 2, we derive an analytic model of conjunctions in the limit of small orbital eccentricities. 
In Section 3, we test our model by comparing it with numerical integrations of the full equations of 
motion, and determine the domain of parameters where the analytic model is valid. In Section 4, 
we show how the magnitude of the conjunction effect scales with different parameters. Finally, 
in Section 5, we discuss how the detection of the conjunction effect can be used to confirm
and characterize transiting planetary systems. 
\section{Analytic Model}
\subsection{Equations of Motion}
We consider a system of two planets with masses $m_1$ and $m_2$ orbiting about a central star with 
mass $M_*$. The planetary orbits are assumed to be nearly coplanar and nearly circular, with planet 
1 on the interior and planet 2 on the exterior. The analytic model is developed in variables 
$a_i$, $\lambda_i$, and $z_i=e_i \exp(\imath \varpi_i)$, where index $i=1,2$ stands for the
two planets. These variables are non-singular for $e_i \rightarrow 0$, and non-canonical, with the later 
being appropriate because we do not aim at developing the theory beyond the lowest order in 
eccentricity and beyond the linear terms in $m_1/M_*$ and $m_2/M_*$. The inclination terms are 
ignored because they appear in the second and higher powers, and are therefore not overly important 
for the TTVs (as we show in Section 4). With that being clarified, the Lagrange equations describing 
the evolution of orbital elements are
\begin{eqnarray}
 \frac{da_i}{dt} & = &  -\frac{2}{\mu_in_ia_i}\frac{\partial {\cal H}_{\rm per}}
  {\partial \lambda_i} \; , \label{p1} \\
 \frac{d\lambda_i}{dt} & = &  n_i+\frac{2}{\mu_in_ia_i}\frac{\partial {\cal H}_{\rm per}}{\partial a_i}
  + {\cal O}(e) \; , \label{p2} \\
 \frac{dz_i}{dt} & = &  -\frac{2\,\imath}{\mu_i n_ia_i^2}\frac{\partial
  {\cal H}_{\rm per}}{\partial z_i} + {\cal O}(e) \; ,
 \label{p3} 
\end{eqnarray} 
where $\mu_i$ is the reduced mass and $n_i$ is the mean motion. All terms that are explicitly the first 
or higher eccentricity powers were removed from Eqs.~(\ref{p2}) and (\ref{p3}). ${\cal H}_{\rm per}$ 
denotes the perturbation part of the Hamiltonian. 

We follow Malhotra (1993) and split the Hamiltonian, ${\cal H}={\cal H}_{\rm Kep}+{\cal H}_{\rm per}$, 
such that  
\begin{equation}
 {\cal H}_{\rm Kep} = -G\,\frac{\mu_1 M_1}{2a_1} - G\,\frac{\mu_2 M_2}{2a_2} \; ,
  \label{hun}
\end{equation}
where $M_1=M_*+m_1$, $\mu_1=M_* m_1/M_1$, $M_2=M_*(M_1+m_2)/M_1$, and $\mu_2=M_1 m_2/(M_1+m_2)$.
This choice implies that the unperturbed motion satisfies $n_1^2a_1^3 = GM_1$ and $n_2^2a_2^3 = 
GM_2$. At the unperturbed level, $a_1$, $a_2$, $z_1$ and $z_2$ are constant, and 
$\lambda_1=n_1t+\lambda_1^{(0)}$ and $\lambda_2=n_2t+\lambda_2^{(0)}$, where $\lambda_1^{(0)}$ 
and $\lambda_2^{(0)}$ are the initial phases at $t=0$.

As for the interaction term, up to the first power in $m_1/M_*$, $m_2/M_*$ and eccentricities, 
we have ${\cal H}_{\rm per} = {\cal H}_0 + {\cal H}_1$, where the part independent of 
eccentricities is  
\begin{equation}
 {\cal H}_0 = G\,\frac{m_1 m_2}{a_2}\left[\alpha\cos\psi-P(\alpha,\psi)\right]\; 
  \label{h0}
\end{equation}
with $\alpha=a_1/a_2$, synodic angle $\psi=\lambda_1-\lambda_2$, and
\begin{equation}
 P(\alpha,\psi) = \left(1 - 2\alpha\cos\psi +\alpha^2\right)^{-1/2}\; . 
  \label{p}
\end{equation}
The first-order eccentricity term, ${\cal H}_1$, is somewhat more complicated:  
\begin{eqnarray}
 {\cal H}_1  =   G\,\frac{m_1 m_2}{a_2} &\times& \Re \biggl(c_0^+\, z_1
  \,e^{-\imath \lambda_1}- d_0^+\, z_2\, e^{-\imath \lambda_2} \label{h1} \\
&+&\sum_{j\geq 1}\left[\left(c_j^-+\alpha\delta_{j1}\right) z_1
  \,e^{-\imath(j\psi+\lambda_1)} + \left(c_j^+-3\alpha\delta_{j1}\right) z_1
  \,e^{\imath(j\psi-\lambda_1)}\right] \nonumber \\
&-&\sum_{j\geq 1}\left[ d_j^- \,z_2\,
  e^{-\imath(j\psi+\lambda_2)} + \left(d_j^+ -4\alpha\delta_{j1}\right) z_2
  \,e^{\imath(j\psi-\lambda_2)}\right]\biggr) \; \nonumber
\end{eqnarray}
(Malhotra 1993, Agol et al. 2005). Here, $\Re$ denotes the real part, $\delta_{j1}$ is the Kronecker delta, 
$c_j^\pm = \partial_\alpha b^{(j)}_{1/2} \pm 2j b^{(j)}_{1/2}$, $d_j^\pm = c_j^\pm +b^{(j)}_{1/2}$,
and $\partial_\alpha b^{(j)}_{1/2} = \alpha\,(\partial/\partial\alpha) b^{(j)}_{1/2}$ for $j\geq 0$.
Symbols $b^{(j)}_{1/2}(\alpha)$ denote the Laplace coefficients. They are best evaluated 
from $b^{(0)}_{1/2}=(2/\pi)\,{\bf K}(\alpha)$ and $b^{(1)}_{1/2}=2\,[{\bf K}(\alpha)-{\bf E}(\alpha)]/(\pi\alpha)$, 
where ${\bf K}$ and ${\bf E}$ are the complete elliptic integrals of the first and second kinds, 
respectively, and the efficient and stable recurrences recommended in Brouwer \& Clemence (1961). 
The derivatives $\partial_\alpha b^{(j)}_{1/2}(\alpha)$ were also obtained from the recurrences 
defined in Brouwer \& Clemence (1961).
\subsection{The First-Order Solution}
The first-order perturbation solution can be obtained by inserting the unperturbed motion (i.e., 
corresponding to ${\cal H}_{\rm Kep}$) into the right-hand sides of the Eqs.~(\ref{p1})-(\ref{p3}), 
and performing a quadrature. Below we compute this quadrature in the time interval from 
$\psi=-\pi$ to $\psi=\pi$, i.e. over one conjunction cycle. For the reasons explained in the next 
paragraph we find it useful to formulate the results of the quadrature in terms of the synodic angle 
$\psi$, rather than of time, but these two formulations are interchangeable because $\psi=(n_1-n_2)t-\pi$ 
(with $\lambda_2^{(0)}-\lambda_1^{(0)}=\pi$). To keep things simple, we perform our calculation 
only to the lowest order in eccentricities, where the resulting expressions become independent of 
$e_1$ and $e_2$. 

If $\alpha$ is somewhat large (but not too large to lead to the co-orbital motion), the orbital spacing 
is relatively tight (as in many Kepler systems), and the interaction between planets happens almost 
exclusively at conjunctions. This is the case when using $\psi$ and the interaction Hamiltonian in Eqs. 
(\ref{h0})-(\ref{h1}) is the most helpful, because the `impulsive' effects of conjunctions are
well captured by a nearly discrete change when $\psi\simeq0$. If, instead, $\alpha$ is small 
($\alpha<0.5$), the conjunction effects cannot be easily isolated, and the Fourier series in Eq. 
(\ref{chi}) becomes more a appropriate representation of the TTVs.  
\subsubsection{Semimajor Axis and Mean Longitude}
We first discuss the variations of semimajor axis and mean longitude described by Eqs.~(\ref{p1})
and (\ref{p2}). In this case, we use ${\cal H}_{\rm per} = {\cal H}_0$ in the right-hand side of 
Eq.~(\ref{p1}), and perform the quadrature to obtain
\begin{eqnarray}
 \delta a_1 & = & -2\,\nu_1 a_1 \,\frac{m_2}{M_*}\,
  Q\left(\alpha,\psi\right)\; , \label{da1} \\
 \delta a_2 & = & 2\,\nu_2 a_2 \,\frac{m_1}{M_*}\,
  Q\left(\alpha,\psi\right)\; , \label{da2}
\end{eqnarray}
where $\delta a_1$ and $\delta a_2$ denote the semimajor-axis variations of the inner and outer 
planets, $\nu_1=n_1/(n_1-n_2)$ and $\nu_2=n_2/(n_1-n_2)$. Assuming that $\delta a_1=\delta a_2 
= 0$ at $\psi=-\pi$ (i.e., $\pi$ away from the conjunction), the function $Q$ can be written as 
\begin{equation}
 Q\left(\alpha,\psi\right) = \alpha\cos\psi - P\left(\alpha,\psi\right) + 1 +
  \frac{\alpha^2}{1+\alpha}\; .  \label{q}
\end{equation}
Here and in the following, $\psi$ has to be understood as an unperturbed angle that linearly increases 
with time ($\psi = (n_1-n_2) t-\pi$, where $n_1$ and $n_2$ are the unperturbed orbital frequencies
defined by ${\cal H}_{\rm Kep}$). Note that Eqs. (\ref{da1}) and (\ref{da2}) obey the law of the 
total angular momentum conservation, because $m_1\,\delta a_1/(n_1 a_1)+m_2\,\delta a_2/(n_2 a_2)=0$.

A change of the semimajor axis leads to a change of the mean motion according to
\begin{equation}
 \delta n_1 = -\frac{3}{2}\frac{n_1}{a_1}\,\delta a_1 \; \label{dn1}\ ,
\end{equation}
and similarly for the outer orbit. This term, together with the derivative in the second term 
in Eq.~(\ref{p2}) (where again ${\cal H}_{\rm per} = {\cal H}_0$), allow us to calculate the 
variation of the mean longitude. The quadrature gives
\begin{equation}
 \delta \lambda_1 = \frac{m_2}{M_*}\,\Bigl[A_1(\alpha)+A_2(\alpha)\,\psi + 
A_3(\alpha)\,\sin\psi + A_4\left(\psi,\alpha\right)\Bigr]\;  \label{dl1}
\end{equation}
with
\begin{eqnarray}
 A_1(\alpha) & \!\!\!=\!\!\! & 3\pi\,\alpha \nu_1^2 \left[1 + \frac{\alpha^2}{1+
  \alpha}\right] \label{cc1c} + 2\,\alpha\nu_1\left[\left(2-3\nu_1\right){\bf K}(\alpha)-
  \frac{2\,{\bf E}(\alpha)}{1-\alpha^2}\right]\; , \nonumber \\
 A_2(\alpha) & \!\!\!=\!\!\! & 3\,\alpha\nu_1^2 \left[1 + \frac{\alpha^2}{1+
  \alpha}\right]\; , \label{a1c}\\
 A_3(\alpha) & \!\!\!=\!\!\! & \alpha^2\nu_1 \left(2+3\nu_1\right)\; ,
  \label{b1c}\\
 A_4(\psi,\alpha) & \!\!\!=\!\!\! & 2\,\alpha\nu_1\left[\left(1-3\nu_1\right)
  \frac{F(\delta,r)}{1+\alpha}-\frac{E(\delta,r)}{1-\alpha}\right]\; , \label{c1c}
\end{eqnarray}
where $F(\delta,r)$ and $E(\delta,r)$ are incomplete elliptic integrals of
the first and second kind with modulus $r=2\sqrt{\alpha}/(1+\alpha)$ and
amplitude $\sin\delta = (1+\alpha)\,P(\alpha,\psi)\,\sin\psi/2$.

Similarly, the mean longitude variation of the outer planet is found to be
\begin{equation}
 \delta \lambda_2 = \frac{m_1}{M_*}\,\Bigl[B_1(\alpha)+B_2(\alpha)\,\psi + B_3(\alpha)\,\sin\psi + 
B_4\left(\psi,\alpha\right)\Bigr]\;  \label{dl2}
\end{equation}
with
\begin{eqnarray}
 B_1(\alpha) & \!\!\!=\!\!\! & -3\pi\,\nu_2^2 \left[1 + \frac{\alpha^2}{1+
  \alpha}\right] \label{cc2c} + 2\,\nu_2\left[3\nu_2\, {\bf K}(\alpha)+
  \frac{2\,{\bf E}(\alpha)}{1-\alpha^2}\right]\; , \nonumber \\
 B_2(\alpha) & \!\!\!=\!\!\! & -3\,\nu_2^2 \left[1 + \frac{\alpha^2}{1+
  \alpha}\right]\; , \label{a2c}\\
 B_3(\alpha) & \!\!\!=\!\!\! & -\alpha\nu_2 \left(4+3\nu_2\right)\; ,
  \label{b2c}\\
 B_4(\psi,\alpha) & \!\!\!=\!\!\! & 2\,\nu_2\left[\left(1+3\nu_2\right)
  \frac{F(\delta,r)}{1+\alpha}+\frac{E(\delta,r)}{1-\alpha}\right]\; . \label{c2c}
\end{eqnarray}

\begin{figure*}
\epsscale{0.6}
\plotone{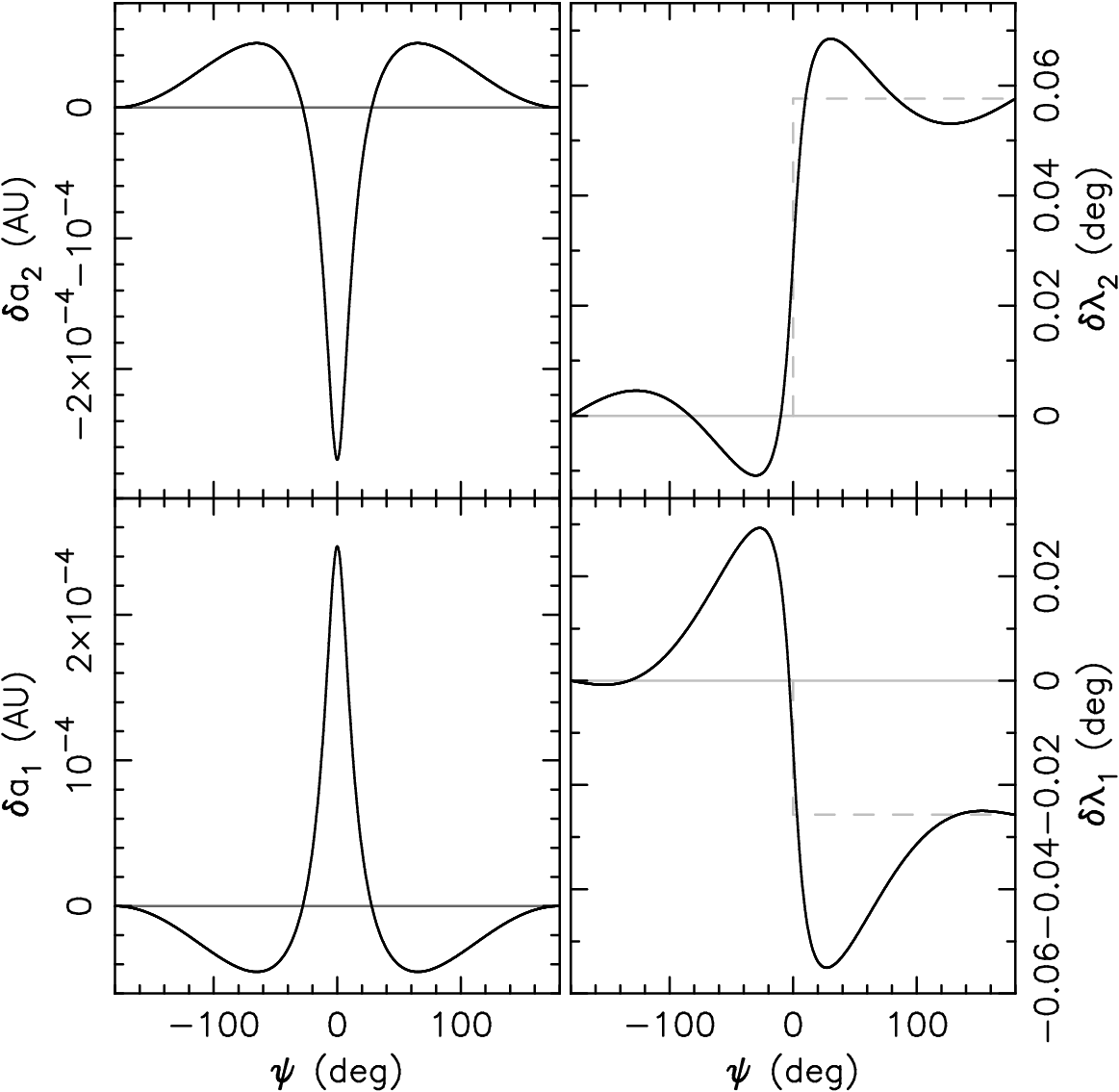}
\caption{The effect of a single conjunction on $\delta a_1$ and $\delta a_2$
(left; computed from Eqs.~(\ref{da1}) and (\ref{da2})), and $\delta \lambda_1$ 
and $\delta \lambda_2$ (right; computed from Eqs.~(\ref{dl1}) and (\ref{dl2})).
Here, the two planets have masses $m_1=m_2=10^{-5}\ M_*$, with $M_*=M_\odot$,
and semimajor axes $a_1=0.84$~AU and $a_2=1$~AU. The evolution is shown as 
a function of the synodic angle $\psi=\lambda_1-\lambda_2$, where $\lambda_1$ 
and $\lambda_2$ are unperturbed mean longitudes of the two planets. The 
conjunction occurs when $\psi \simeq 0$. Notably, the semimajor axis of the
inner (outer) planet increases (decreases) during the conjunction, while the 
mean longitudes of the two planets suffer a step-like discontinuity.}
 \label{f1}
\end{figure*}

The expressions in Eqs.~(\ref{dl1}) and (\ref{dl2}) have the same structure revealing 
the underlying effects of planetary conjunctions.  The first term in the right-hand 
sides of these equations, independent of $\psi$, reflects a nearly step-like discontinuity 
of the mean longitude near the conjunction. The amplitude of this discontinuity is 
equal to $2 A_1 m_2/M_*$ for the inner planet, and $2B_1 m_1/M_*$ for the outer planet.
Figure~\ref{f1} shows an example for $a_1=0.84$~AU, $a_2=1$~AU, $m_1/M_*= m_2/M_*=10^{-5}$,
and $M_*=M_\odot$, where $M_\odot$ is the solar mass. 
\subsubsection{Eccentricity and Apsidal Longitude}
Equation (\ref{deltat}) shows that the TTVs depend not only on the variation of the mean 
longitude, but also on variation of eccentricity and apsidal longitude. To the lowest 
order in eccentricity, $\delta t_i$ becomes
\begin{equation}
-n_i\,\delta t_i = \delta \lambda_i + \delta \lambda_i^{\rm eff} + {\cal O}(e)\; ,  \label{dt}
\end{equation}
where $\delta \lambda_i$ was computed in Section 2.2.1, and $\delta \lambda_i^{\rm eff}$ is 
the effective contribution from the short-periodic variations of $e_i$ and $\varpi_i$
(e.g., Nesvorn\'y \& Morbidelli 2008). Using the complex variable $z_i$ defined in Section 2.1. 
we have 
\begin{equation}
 \delta \lambda_i^{\rm eff} = \imath\left(\delta z_i\, e^{-\imath
  \lambda_i} - \delta z_i^*\, e^{\imath \lambda_i} \right)
  \; , \label{lameff}
\end{equation}
where $\delta z_i$ is the variation of $z_i$, $\delta z_i^*$ is the complex conjugate,
and $\lambda_i$ is the unperturbed mean longitude. If the actual transits occur when 
$\lambda_i\simeq0$, the above expression reduces, consistently with Eq. (\ref{deltat}), to 
$\delta \lambda_i^{\rm eff}=-2\delta h_i+{\cal O}(e_i)$.

The first-order approximation of $\delta z_i$ is obtained by inserting unperturbed motion
into the right-hand side of Eq.~(\ref{p3}) and performing the quadrature. We find that
\begin{equation}
 \delta \lambda_1^{\rm eff} = \frac{m_2}{M_*}\Bigl[C_1(\alpha)
  \sin\left[\nu_1\left(\psi+\pi\right)\right] + \sum_{j\geq 1} D_1^j(\alpha)
  \sin j\psi \Bigr]  \; , \nonumber  \label{dl1eff} 
\end{equation}
where 
\begin{eqnarray}
C_1(\alpha) & \!\!\!=\!\!\! & \alpha \biggl[ \partial_\alpha b_{1/2}^{(0)}
  + 2\,\alpha\,\frac{\nu_1\left(3\nu_1-2\nu_2\right)}{\nu_2\left(2\nu_1-
  \nu_2\right)} -2\,\nu_1 \sum_{j\geq 1} \left(-1\right)^j
  \frac{\nu_1\,\partial_\alpha b_{1/2}^{(j)}+2j^2\,b_{1/2}^{(j)}}{j^2-\nu_1^2}
  \biggr]\; , \label{aa1c} \\
D_1^j(\alpha) & \!\!\!=\!\!\! & 2\,\delta_{j1}\,\alpha^2 \frac{\nu_1\left(3\nu_1-2\nu_2\right)}
  {\nu_2\left(2\nu_1-\nu_2\right)} +
2j\,\alpha\nu_1\frac{\partial_\alpha
  b_{1/2}^{(j)}+2\nu_1\,b_{1/2}^{(j)}}{j^2-\nu_1^2}\; .  \label{bb1jc} 
\end{eqnarray}
Similarly, for the outer planet we obtain
\begin{equation}
 \delta \lambda_2^{\rm eff} = \frac{m_1}{M_*}\Bigl[C_2(\alpha)
  \sin\left[\nu_2\left(\psi+\pi\right)\right] + \sum_{j\geq 1} D_2^j(\alpha)
  \sin j\psi \Bigr] \; ,  \label{dl2eff}
\end{equation}
where
\begin{eqnarray}
\!\!\!\!\!\!  C_2(\alpha) & \!\!\!=\!\!\! & 4\,\alpha\,\frac{\nu_2}{\left(\nu_1-
  2\nu_2\right)} - \left(\partial_\alpha b_{1/2}^{(0)} +b_{1/2}^{(0)}
  \right) +2\,\nu_2 \sum_{j\geq 1} \left(-1\right)^j
  \frac{\nu_2\, \partial_\alpha b_{1/2}^{(j)}+\left(2j^2+\nu_2\right)
  b_{1/2}^{(j)}}{j^2-\nu_2^2}\biggr]\; , \label{aa2c} \\
\!\!\!\!\!\!  D_2^j(\alpha) &\!\!\! =\!\!\! & 4\,\delta_{j1}\,\alpha \frac{\nu_2}{\nu_1-2\nu_2}
-2j\,\nu_2\frac{\partial_\alpha
  b_{1/2}^{(j)}+\left(2\nu_2+1\right)\,b_{1/2}^{(j)}}{j^2-\nu_2^2} \label{bb2jc} \; .
\end{eqnarray}

The coefficients $C_1(\alpha)$, $C_2(\alpha)$, $D_1^j(\alpha)$ and $D_2^j(\alpha)$
defined above have singularities at the first-order mean motion resonances, which is 
a consequence of the perturbation method applied here. 
For example, when $n_1/n_2=j/(j-1)$, we have $\nu_1=j$, which causes a zero divisor 
in Eqs. (\ref{aa1c}) and (\ref{bb1jc}). Similarly, when $n_1/n_2=(j+1)/j$ we have 
$\nu_2=j$, implying zero divisors in Eqs. (\ref{aa2c}) and (\ref{bb2jc}). 
Interestingly, however, when these expressions combine in Eqs. (\ref{dl1eff})
and (\ref{dl2eff}) into variables important for the TTVs, the singularities
disappear such that both $\delta \lambda_1^{\rm eff}$ and $\delta \lambda_2^{\rm eff}$
are well defined at resonances. This can be most easily verified by assuming
that $n_1/n_2=j/(j-1)+\varepsilon$ or $n_1/n_2=(j+1)/j+\varepsilon$, 
where $\varepsilon$ is a small quantity, and showing that $\lambda_1^{\rm eff}$
and $\lambda_2^{\rm eff}$ are non-divergent when $\varepsilon\rightarrow 0$.

To derive the expressions in Eqs. (\ref{dl1eff}) and (\ref{dl2eff}) we assumed 
that $e_1=0$ and $e_2=0$ when $\psi=-\pi$. Note that, in this case, $\delta \lambda_i^{\rm eff}$ 
is independent of the initial phases $\lambda_1^{(0)}$ and $\lambda_2^{(0)}$. Together
with a similar result obtained in Section 2.2.1, this implies that $\delta t_i$
will also be independent of $\lambda_i^{(0)}$ (see Section 2.3). It will only depend 
on the orbital period, $m_1/M_*$, $m_2/M_*$ and $\alpha$.

\begin{figure*}
\epsscale{0.6}
\plotone{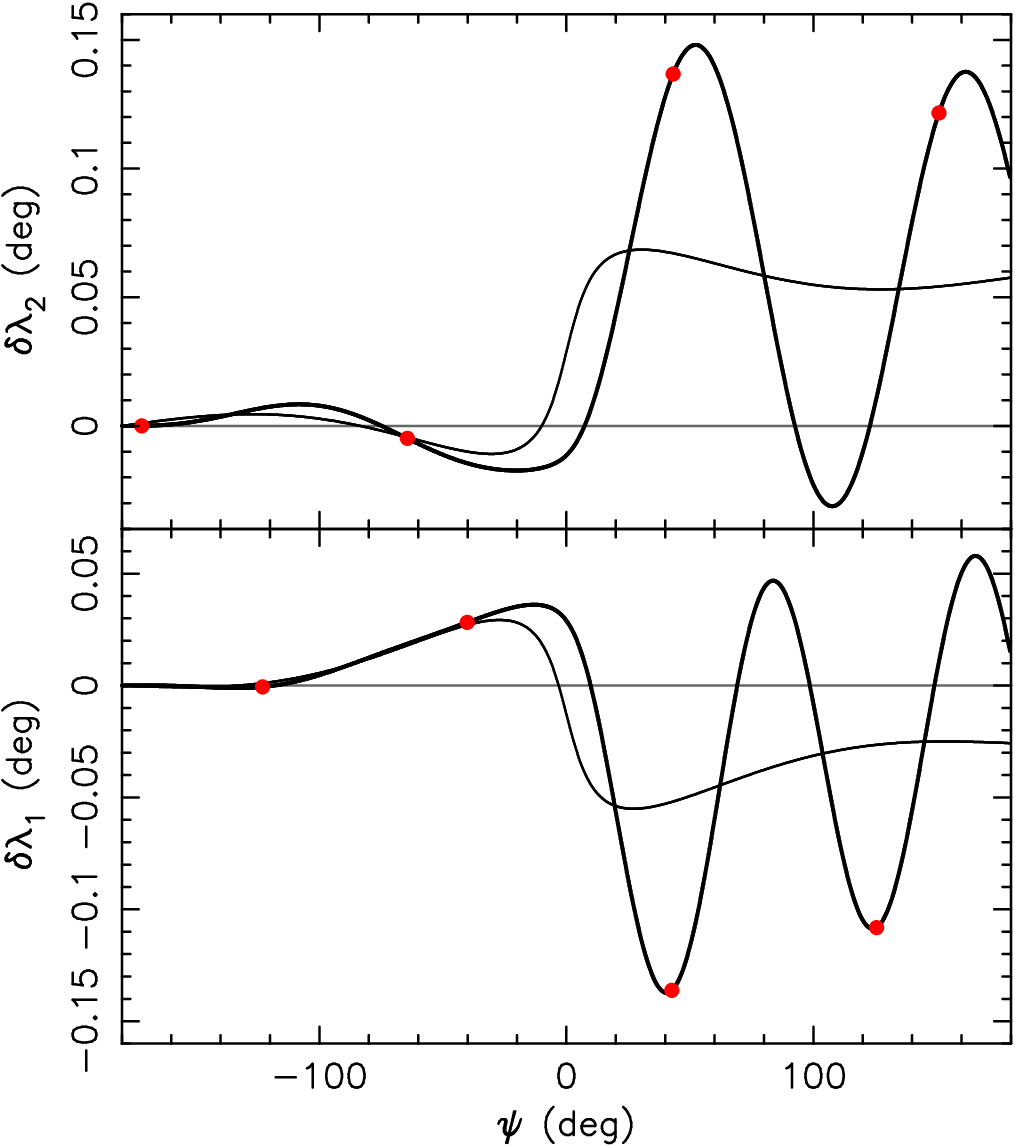}
\caption{The effect of a single conjunction on $\delta \lambda_1$ (bottom panel)
and $\delta \lambda_2$ (top panel) for two planets with masses $m_1=m_2=10^{-5}\ 
M_*$ ($M_*=M_\odot$), semimajor axes $a_1=0.84$~AU and $a_2=1$~AU, and initially
circular orbits. The conjunction between the two planets occurs when $\psi = 
\lambda_1-\lambda_2 \simeq 0$. The thin lines show the direct variation of 
$\delta \lambda_i$ from Eqs.~(\ref{dl1}) and (\ref{dl2}). The bold lines show
the combined effect, $\delta \lambda_i + \delta \lambda_i^{\rm eff}$, where 
$\delta \lambda_i^{\rm eff}$ stands for the effective contribution of eccentricities 
and apsidal longitudes. Assuming that both planets are transiting the red symbols 
show the expected transit cadence. }
\label{f2}
\end{figure*}

Figure~\ref{f2} illustrates the role of $\delta \lambda_i^{\rm eff}$ in an example with 
$a_1=0.84$~AU, $a_2=1$~AU, $m_i=10^{-5}\ M_*$, $M_*=M_\odot$, and zero initial 
eccentricities. The effects of $\delta \lambda_i^{\rm eff}$ are initially small, but 
when small orbital eccentricities are excited during the conjunction (i.e., 
when $\psi\simeq0$), $\delta t_i$ becomes a composite of two terms with comparable
magnitudes: (i) a step-like change produced by the direct variation $\delta \lambda_i$ 
(Section 2.2.1), and (ii) oscillations from $\delta \lambda_i^{\rm eff}$.
According to Eqs. (\ref{dl1eff}) and (\ref{dl2eff}), the oscillatory part has 
a rich spectrum of Fourier terms with frequencies $n_1$, $n_2$ and $j(n_1-n_2)$.

A general solution corresponding to small values of the initial eccentricities
$e_1^{(0)}$ and $e_2^{(0)}$, can be obtained by adding $2\,e_1^{(0)} \sin\left[\nu_1\left(\psi
+\phi_1^{(0)}\right)\right]$ to Eq. (\ref{dl1eff}) and $2\,e_2^{(0)} \sin\left[\nu_2\left(\psi
+\phi^{(0)}_2\right)\right]$ to Eq. (\ref{dl2eff}). Here, 
$\phi_i^{(0)}=(\lambda_i^{(0)}-\varpi_i^{(0)})/\nu_i+\pi$, where $\varpi_i^{(0)}$ are
the apsidal longitudes of the two orbits when $\psi=-\pi$. Since, by definition,
$\lambda_2^{(0)}=\lambda_1^{(0)}+\pi$, the general solution adds $e_i^{(0)}$, 
$\varpi_i^{(0)}$ and $\lambda_1^{(0)}$ (or $\lambda_2^{(0)}$) to the list of parameters.
\subsection{Transit Timing Variations}
The equations derived above can be used to compute $\delta \lambda_i + \delta 
\lambda_i^{\rm eff}$ as a function of the synodic angle $\psi$ in the interval 
$-\pi \leq \psi < \pi$. To be able to track changes over the successive rotations of $\psi$, 
we need to add a constant term to $\delta \lambda_i$ that expresses how $\delta \lambda_i$ 
changed during the previous conjunction. For an arbitrary $\psi \geq -\pi$, and denoting 
${\rm {\bf int}}[x]$ the integer part of $x$, this term is
\begin{equation}
\delta \lambda_1^{\rm add} = 2\,\frac{m_2}{M_*}\,A_1(\alpha)\,
{\rm {\bf int}}\left[(\psi+\pi)/2\pi\right]\;   \label{dl1add}
\end{equation}
for the inner planet, and
\begin{equation}
\delta \lambda_2^{\rm add} = 2\,\frac{m_1}{M_*}\,B_1(\alpha)\,
{\rm {\bf int}}\left[(\psi+\pi)/2\pi\right]\;   \label{dl2add}
\end{equation}
for the outer planet. 
Adding this term successively each time $\psi$ increases by $2\pi$ leads to a 
situation where $\delta t_i$ either decreases (for the inner planet) or increases 
(for the outer one) in a series of steps. 
Equations (\ref{dl1eff})-(\ref{bb2jc}) express the general dependence of $\delta 
\lambda_i^{\rm eff}$ on $\psi$ in that they can be used to compute the 
conjunction effect over the successive rotations of $\psi$. To do so,
$\psi=(n_1-n_2)t-\pi$ needs to be substituted into Eqs. (\ref{dl1eff}) and (\ref{dl2eff}).
With these provisions, $\delta t_i$ can be computed for any $t \geq 0$. 

Here we are mainly interested in the transit timing. We therefore
assume that the observer detects transits on a (nearly) linear ephemeris, $t_i^k= k P_i+t_i^0$, 
where $t_i^0>0$ is the epoch of the first transit, $k\geq 1$ denotes the subsequent 
transit cycles, and $t_i^k$ are the subsequent transit epochs. This gives $\psi_i^k=2 \pi 
k/\nu_i + (n_1-n_2) t_i^0 -\pi$, where $\psi_i^k$ denotes the value of the synodic angle 
corresponding to the $k+1$ transit of planet $i$. Substituting this into Eqs.
(\ref{dl1eff}) and (\ref{dl2eff}), and dropping all constant terms (i.e., those independent 
of $k$), we find that:
\begin{equation}
\delta t_1^k = -{P_1 \over 2\pi} {m_2 \over M_*} \left[ 
A(\psi_1^k) 
+ 2 A_1{\rm {\bf int}}\left[ {k \over \nu_1} + {t_1^0 \over P_\psi} \right] 
+\sum_{j\geq 1} (-1)^j D_1^j \sin \left( 2 \pi j \left[{k \over \nu_1} + {t_1^0 \over P_\psi}
\right] \right) \right]  \label{dt1}
\end{equation}
and 
\begin{equation}
\delta t_2^k = -{P_2 \over 2\pi} {m_1 \over M_*} \left[ 
B(\psi_2^k) 
+ 2 B_1{\rm {\bf int}}\left[ {k \over \nu_2} + {t_2^0 \over P_\psi} \right] 
+\sum_{j\geq 1} (-1)^j D_2^j \sin \left( 2 \pi j \left[{k \over \nu_2} + {t_2^0 \over P_\psi}
\right] \right) \right]  \label{dt2}
\end{equation}
Here we denoted $A(\psi_1^k)=A_2 \psi_1^k+A_3\sin \psi_1^k + A_4(\psi_1^k)$,
$B(\psi_2^k)=B_2 \psi_2^k+B_3\sin \psi_2^k + B_4(\psi_2^k)$, and synodic period 
$P_\psi=2 \pi/(n_1-n_2)$. Note that $A$, $B$, $A_1$, $B_1$ $D_1^j$ and $D_2^j$ 
are explicit functions of $\alpha$ as defined in Sections 2.2.1 and 2.2.2. Also
note that Eqs. (\ref{dt1}) and (\ref{dt2}) are related to Eqs. (A7) and (A8) 
previously derived in the appendix of Agol et al. (2005).

Equations (\ref{dt1}) and (\ref{dt2}) describe the effect of conjunctions on the TTVs.
The three terms present in these equations stand for: the (i) modulation of $\delta t_i$ 
from $\delta \lambda_i$ during a single rotation of the synodic angle, (ii) total 
change of $\delta t_i$ from $\delta \lambda_i$ over all previous synodic cycles, and (iii) 
periodic terms from the eccentricity-related perturbations. The expressions for 
(iii) are non-divergent, because if $\nu_1=j$ (or $\nu_2=j$) the variable part of the 
corresponding sinusoidal terms vanishes from Eq. (\ref{dt1}) (or Eq. (\ref{dt2})). 

Notably, the sinusoidal terms in Eqs. (\ref{dt1}) and (\ref{dt2}) originate from terms 
with $\sin j \psi$ (see  Eqs. (\ref{dl1eff}) and (\ref{dl2eff})) and are $2\pi$-periodic 
in $\psi$. Therefore, as far as these periodic terms are concerned, there is no
difference between the first, second, or any other cycle of $\psi$. The difference 
between the TTVs in different cycles of $\psi$ arises, instead, by how fixed periodic 
terms are sampled by TTV observations. For given $\alpha$, it turns out that the periodic 
term with the largest amplitude $D_1^j(\alpha)$ (or $D_2^j(\alpha)$) is the one with 
$j \sim \nu_1$ in Eq. (\ref{dt1}) (or $j \sim \nu_2$ in Eq. (\ref{dt2})). As shown
in Eq. (\ref{dt1}) (or Eq. (\ref{dt2}), however, these terms will be sampled with
$\sim 2 \pi k$ cadence and will therefore have only a limited impact on the TTVs.    
\section{Validity Domain of the Analytic Model}
The analytic model developed in the previous sections can be used as a guideline 
to understand the basic effects of conjunctions on the TTVs. We will discuss these 
effects, and their scaling with different parameters, in Section 4. Before we do 
so, however, we first establish the domain of validity of the analytic model 
by comparing the results to those obtained from an exact $N$-body integration. This 
comparison was done with the $N$-body code described in Nesvorn\'y et al. (2013), where 
the TTVs are computed with an efficient and precise algorithm (also see Deck et al. 2014).  

Figure \ref{comp1} illustrates the results for the test case previously 
shown in Figures~\ref{f1} and \ref{f2}. Here the results of the analytic model are in 
an excellent agreement with those obtained from the $N$-body integration. This was expected 
because the parameters of the test case were set to be in the domain where the 
analytic model should be valid (e.g., small planetary masses, $\alpha$ not too large, 
and $e_i^{(0)}=0$).
 
In general, however, the analytic model is obviously only an approximation of the 
conjunction effect. First, the two orbits were assumed to be strictly coplanar.
Second, we assumed that $m_i \ll M_*$, such that ${\cal H}_{\rm per}$ can be treated as a 
perturbation of ${\cal H}_{\rm Kep}$. Terms of the second and higher orders in $m_i/M_*$ were 
not included. Third, we assumed that $e_i \ll 1$, expanded the Hamiltonian in powers
of $e_i$, and retained only the lowest power of $e_i$. We therefore expect the 
analytic model to be valid only for very nearly circular orbits of both planets. 

Figure \ref{limit1} illustrates the approximate nature of the analytic model. To make 
this figure, we surveyed a range of orbital separations and eccentricities ($0.5<\alpha<0.9$,
$0<e_1^{(0)}<0.1$ and $0<e_2^{(0)}<0.1$). Other parameters were held fixed ($m_i=10^{-5}\, M_*$, 
$M_*=M_\odot$). In each case, we followed dynamics over one conjunction cycle and determined 
the (i) amplitude of $\delta t_i$ variation produced as a result of conjunction, and 
(ii) the difference between analytically and numerically computed $\delta t_i$ when 
$\psi$ approaches $\pi$ (i.e., at the end of the conjunction cycle). From this, by dividing 
(ii) by (i), we computed the relative error of the analytic model as a function of $\alpha$, 
$e_1^{(0)}$ and $e_2^{(0)}$. 

From Figure \ref{limit1} we see that the analytic model is valid only for small eccentricities, 
and the eccentricity threshold ($e_{\rm crit}$) beyond which the relative error is excessive 
(say $>$10\%) is a strong function of the radial separation between orbits. For example, 
$e_{\rm crit}\simeq 0.1$ for $\alpha=0.5$ while $e_{\rm crit}\simeq 0.01$ for $\alpha=0.8$,
This is expected because the second and higher order effects in $m_i/M_*$, neglected in our 
analytic model, should become important with increasing $\alpha$. Also, we would need to 
include eccentricity terms beyond the lowest power to make the analytic model more generally 
valid for larger eccentricities.  
   
We used $m_i/M_*=10^{-5}$ in Figure \ref{limit1}, but it turns out that the general 
appearance of this figure is independent of the considered planetary masses. This is because
both the magnitude of the conjunction effect and the error of the analytic model 
increase (nearly) linearly with $m_i/M_*$. The relative error therefore remains approximately 
the same. 

The mass ratio $m_i/M_*$ sets the limit in $\alpha$ beyond which the analytic model
does not apply. Beyond this limit, the co-orbital dynamics appears and the two planets 
can switch positions radially (i.e., following the horseshoe or tadpole trajectory the 
inner planet becomes an outer one, and vice versa). In this situation,
$\alpha$, as defined here, evolves from $\alpha<1$ to $\alpha>1$, and the Fourier 
expansion of ${\cal H}_{\rm per}$ in Section 2.1 becomes divergent. The TTVs occurring 
for two planets in the co-orbital regime were recently investigated by Vokrouhlick\'y 
\& Nesvorn\'y (2014). 

Finally, as we already mentioned at the beginning of Section 2.2, our analytic model is valid 
but not really useful if $\alpha$ is small. In such a case, the conjunctions between planets
cannot be described as a discrete effect, because the gravitational interaction of planets is 
similarly strong for any phase of $\psi$. For small $\alpha$, we therefore find it more intuitive 
to use the representation in Eq. (\ref{chi}), where the TTVs are fully expanded in the Fourier 
series. The transition between the two regimes is gradual such that it is difficult to establish 
a single value of $\alpha$ where this transition happens. We roughly find that our analytic model 
of conjunctions is useful for $\alpha > 0.5$, while the Fourier series representation becomes 
more adequate for $\alpha < 0.5$.
\section{Scaling of the Conjunction Effect with Planetary Properties}
According to Eqs. (\ref{dt1}) and (\ref{dt2}), the expected variation of transit timing, $\delta t_i$, 
is proportional to $P_i$, where $P_i$ is the orbital period of the transiting planet. Thus, assuming that 
the observation baseline is long enough to cover several conjunction cycles, the detection of the 
conjunction effect would be easier for planets with longer orbital periods, for which the effect 
is larger. In reality, however, the current observational baselines are typically only a few years such 
that we do not expect that the TTVs to be generally detectable for long-period planets. The conjunction 
effect could potentially be detected for long-period planets only if at least a few transits were observed 
before the conjunction and a few transits after the conjunction, which would require a fortuitous 
configuration of the planetary system at the current epoch (a good example of this can be KOI-351g; 
Cabrera et al. 2014).  

The scaling of $\delta t_i$ with the planetary and stellar masses is obvious from Eqs. (\ref{dt1}) and 
(\ref{dt2}), at least in the approximation $m_i \ll M_*$ that we adopted in the analytic model. While 
$\delta t_1$ scales linearly with $m_2/M_*$, $\delta t_2$ scales linearly with $m_1/M_*$. This means
that a detection of the conjunction effect in transits of the inner planet can help to determine 
the mass of the outer planet, and vice versa. Also, the detection of transit variations is obviously 
easier in a system with more massive planets and, for a fixed orbital period, with lower stellar 
mass.\footnote{Note that if more than two planets are present in a given system, the TTVs from 
conjunctions of different pairs should add linearly, at least in the approximation of our analytic model.}

Figure \ref{scale1} illustrates the dependence of $\delta t_i$ on $\alpha=a_1/a_2$. The dashed lines 
in the figure show the amplitude of the $\delta \lambda_i/n_i$ change from a single conjunction between
planets. From Eqs. (\ref{dl1}) and (\ref{dl2}), the amplitude is $2A_1(\alpha) m_2/M_*$ for the inner 
planet and $2 B_1(\alpha) m_1/M_*$ for the outer planet, where $A_1(\alpha)$ and $B_1(\alpha)$ are given 
in terms of the complete elliptic integrals in Eqs. (\ref{cc1c}) and (\ref{cc2c}). 
The total conjunction effect from $\delta\lambda_i+\delta\lambda_i^{\rm eff}$ is shown by solid lines 
in Figure \ref{scale1}. The basic tendency is that the magnitude of the conjunction effect strongly 
increases with $\alpha$, such that it is $\simeq100$-200 times stronger for planets with $\alpha\simeq0.9$
than for planets with $\alpha\simeq0.5$. This is reasonable because the closely packed planetary  
systems are expected to have stronger gravitational interactions. 

In the example given in Figure \ref{scale1} with $m_1=m_2=10^{-5}\ M_*$, $M_*=M_\odot$ and $a_2=1$~AU, 
the magnitude of $\delta t_i$ ranges from 3 minutes (inner planet, $\alpha=0.5$) to over 10 hours (outer 
planet, $\alpha=0.9$). Assuming instead that the outer planet with $a_2=1$ AU has one Earth mass, the 
TTVs of the inner planet should range between $\simeq1$ minute and $\simeq3$ hours. They should 
therefore be generally detectable with adequate photometric precision.    

Figure \ref{scale2} shows how the amplitude of the conjunction effect changes with eccentricity. Given 
that our analytic model loses precision with increasing eccentricity (see Figure 5), here we used our 
$N$-body code to compute $\delta t_i$ over one conjunction cycle. In addition to changing $\alpha$ as 
in Figure \ref{scale1}, we also varied the initial eccentricities of the two planets. Figure \ref{scale2}
shows that the amplitude of the conjunction effect can increase by a factor of $\simeq2$-10 by increasing
the eccentricity from 0 to 0.1. This is significant, because it shows that the likelihood of detection 
of the conjunction effect can be boosted for orbits with modest eccentricities, a case that should 
presumably be common among planetary systems. 

The magnitude of the conjunction effect for eccentric orbits, however, also depends on the relative 
orientation of orbits, as given by $\varpi_1$ and $\varpi_2$, and on where exactly the conjunction
happens along the orbits. The magnitude can increase or decrease, roughly reflecting the physical distance 
between planets during the conjunction. Figure \ref{scale2} was produced by surveying all orbital 
configurations, $0 \leq \varpi_1<2\pi$ and $0 \leq \varpi_2<2\pi$, and plotting the one for which 
the magnitude was maximal. 

Figure \ref{scale3} shows how the amplitude of the conjunction effect changes with the mutual inclination
between orbits, $i_{\rm mutual}$. As in Figure \ref{scale2} we used the $N$-body code to compute 
$\delta t_i$ over one conjunction cycle. Figure \ref{scale2} illustrates that the amplitude of the conjunction 
effect is not very sensitive to $i_{\rm mutual}$. The magnitude varies only up to $\simeq$20\% for 
$i_{\rm mutual}<50^\circ$, relative to to the case with $i_{\rm mutual}=0$. See Nesvorn\'y et al. (2009) 
for a more general analysis of the dependence of the short-periodic TTVs on orbital inclinations.
\section{Discussion}\label{concl}
Equations (\ref{dt1}) and (\ref{dt2}) express our expectation for the TTVs produced by two planets 
on nearly circular and coplanar orbits. The first two terms in these equations result from the 
direct perturbation of the mean longitude. If $\alpha$ is sufficiently close to 1, these terms are
essentially equivalent to a succession of transit timing discontinuities occurring at orbital conjunctions
between planets. The amplitude of these discontinuities is $2A_1(\alpha) m_2/M_*$ for the inner planet
and $2 B_1(\alpha) m_1/M_*$ for the outer planet, where $A_1(\alpha)$ and $B_1(\alpha)$ were defined 
in Eqs. (\ref{cc1c}) and (\ref{cc2c}) and were illustrated in Figure \ref{scale1} (dashed lines). 
The transit times of the inner planet are expected to be delayed relative to a fixed Keplerian ephemeris, 
while those of the outer planet are expected to be sped up. 

The long-term effects of conjunctions, with $\delta t_i$ steadily accumulating over many periods of 
the synodic angle, can be absorbed by a small change of the orbital period.  
The short-period effects of conjunctions, frequently described as `chopping' of the TTV signal 
(e.g, Holman et al. 2010, Carter et al. 2012), should not be mistaken with anything else. When 
the long-term conjunction effects are removed from the transit ephemeris, the TTV signal
from direct perturbation of the mean longitude should have a saw-like profile with each tooth
being marked by a few rising and a few declining transits. If such a saw-tooth profile is identified 
in the data, the mass of planetary companion can be extracted from these measurements,
assuming that $\alpha$ is known, by using Eqs. (\ref{dt1}) or (\ref{dt2}). If $\alpha$
is unknown, Eqs. (\ref{dt1}) or (\ref{dt2}) can be used to constrain $A_1(\alpha)m_2/M_*$ or 
$B_1(\alpha)m_1/M_*$.    

As an example, we discuss the chopping in the TTVs of KOI-884.02 and Kepler-36c (Figure 1).
As for Kepler-36, $P_2\simeq16.2$ day, $\alpha\simeq0.9$, $M_1/M_*\simeq1.3\times10^{-5}$ from 
Carter et al. (2012). The size of the conjunction step computed from Eq. (34) for these 
parameters is $\delta t_2 \sim 0.3$ hour. For a comparison, Figure 1b shows that the actual 
steps during conjunctions are smaller, but not much smaller, than one hour. As for 
KOI-884.02, $P_1\simeq20.5$~day, $\alpha\simeq0.34$, $M_2/M_*\simeq3\times10^{-3}$ from 
Nesvorn\'y et al. (2012). We compute that $\delta t_1 \sim 0.9$ hour from Eq. (33).
For a comparison, the actual steps in KOI-884.02's TTVs, best seen for transit cycles between 
32 and 45 (Figure 1a), are $\sim$1-2 hours. The difference is probably caused by small but 
significant orbital eccentricities (Nesvorn\'y et al. 2014).

\begin{figure*}
\epsscale{0.6}
\plotone{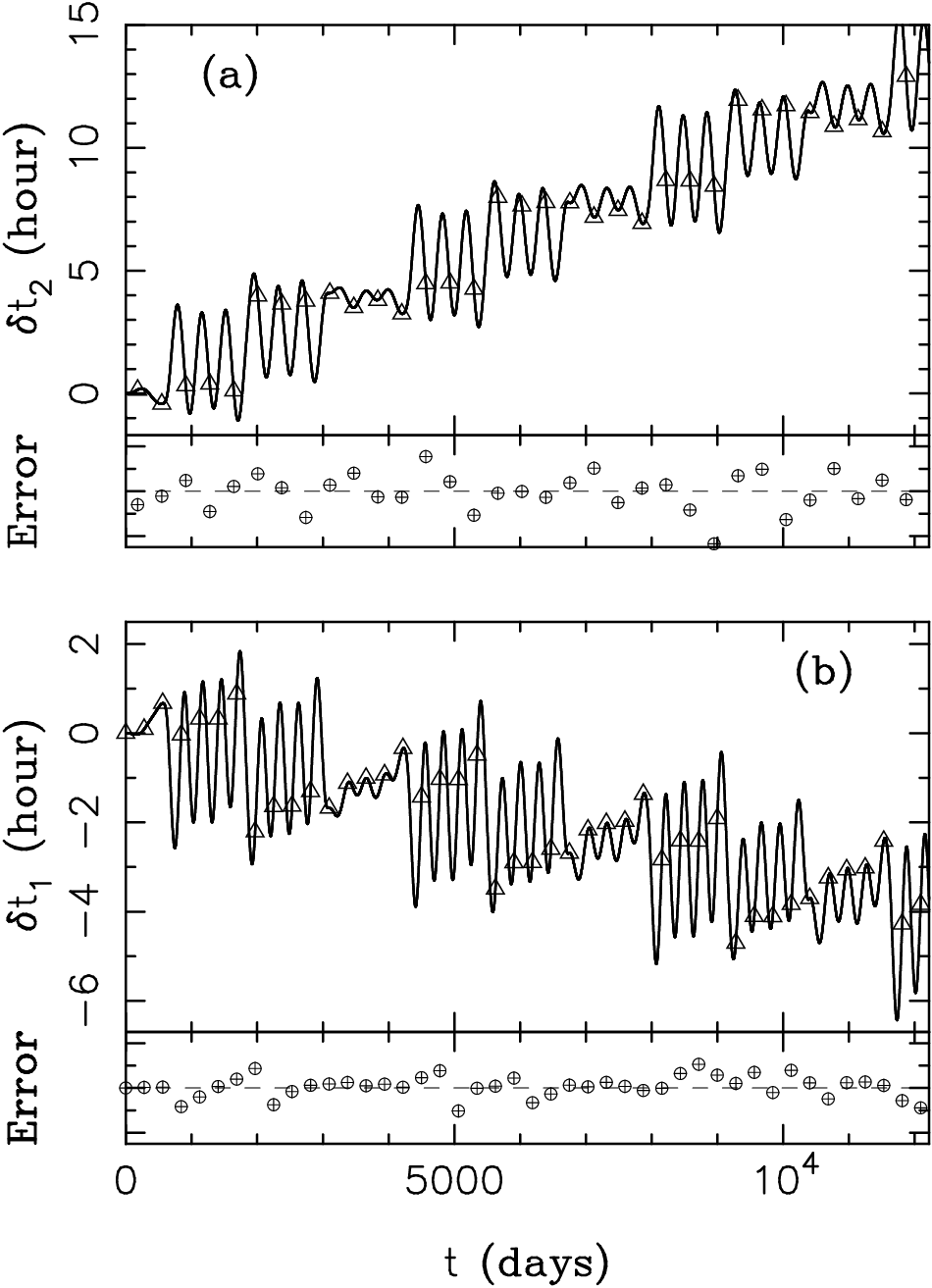}
\caption{A comparison of the analytic model with numerical integration. The triangles 
show the actual TTVs obtained from a numerical integration. The solid lines show  
$\delta t_i$ as a function of time, where $\delta t_i$ were computed from 
the analytic model described in Section 2. The differences between analytic and numeric 
times are shown in the bottom of each panel. These values have to be multiplied by a factor 
of 60 to appear on the same scale with the upper plots (the dashed horizontal line
shows zero for a reference). The two planets have masses $m_1=m_2=10^{-5}\ M_*$ 
($M_*=M_\odot$), semimajor axes $a_1=0.84$~AU and $a_2=1$~AU (thus $P_2/P_1\simeq1.3$), 
and $e_i^{(0)}=0$. The initial mean longitudes were chosen such that $\psi=-\pi$.}
 \label{comp1}
\end{figure*}

The oscillatory part of Eqs. (\ref{dt1}) or (\ref{dt2}) offers a different method to constrain
planetary masses and/or $\alpha$. Here it can be useful to perform the Fourier analysis of
$\delta t_i$. The expectation is that this will reveal frequencies that are integer 
multiples of $(n_1-n_2)$. Some of these frequencies will be faster than the Nyquist frequency,
$f^{\rm N}_i=\pi/P_i$, and will be aliased to the part of the Fourier spectrum with $f<f^{\rm N}_i$.
For example, in the test case shown in Figure \ref{comp1}, $f^{\rm N}_1=0.01117$ d$^{-1}$.
Therefore, frequencies $(n_1-n_2)=0.00514$ d$^{-1}$ and $2(n_1-n_2)=0.01028$~d$^{-1}$ appear
unaliased, while all $j(n_1-n_2)$ frequencies with $j\geq 3$ are aliased to $2 f^{\rm N}_1-j(n_1-n_2)$. 
For example, $3(n_1-n_2)=0.01542$ d$^{-1}$ appears at $0.006933$ d$^{-1}$.

As for KOI-884, $(n_1-n_2)=0.201$ d$^{-1}$ and $f^{\rm N}_1=0.153$ d$^{-1}$. As $(n_1-n_2)>f^{\rm N}_1$ 
in this case, the synodic frequency and all its multiples will be aliased. For example, 
$(n_1-n_2)$ should appear at $0.105$ d$^{-1}$, and $2(n_1-n_2)$ should appear at $0.096$ d$^{-1}$.
The Fourier analysis of the best fit TTV model from Nesvorn\'y et al. (2014) confirms this.
It shows that the peak power density of the $2(n_1-n_2)$ term is about five times larger than
that of the $(n_1-n_2)$ term, as expected from Eqs. (27) and (33). Unfortunately, these terms
are much harder to identify in the existing TTV data of KOI-884.02, because of the short 
coverage, gaps, measurement errors, and other issues. We have done a similar analysis for 
Kepler-36, but do not discuss it here, except for pointing out that the aliasing is not 
a problem for this system, because the synodic frequency $(n_1-n_2)$ is relatively slow 
(as $\alpha\simeq0.9$).  

Identifying the structure of unaliased and aliased frequencies in the frequency domain can be 
useful for the interpretation of the TTV observations. As the amplitudes of these terms are 
proportional to $D_i^j$, as shown in Eqs. (\ref{dt1}) or (\ref{dt2}), the TTV measurements 
can be potentially inverted to obtain a unique determination of the planetary mass and 
orbital separation (e.g., Nesvorn\'y et al. 2013). This highlights the importance of the 
conjunction effect. 

\begin{figure*}
\epsscale{0.6}
\plotone{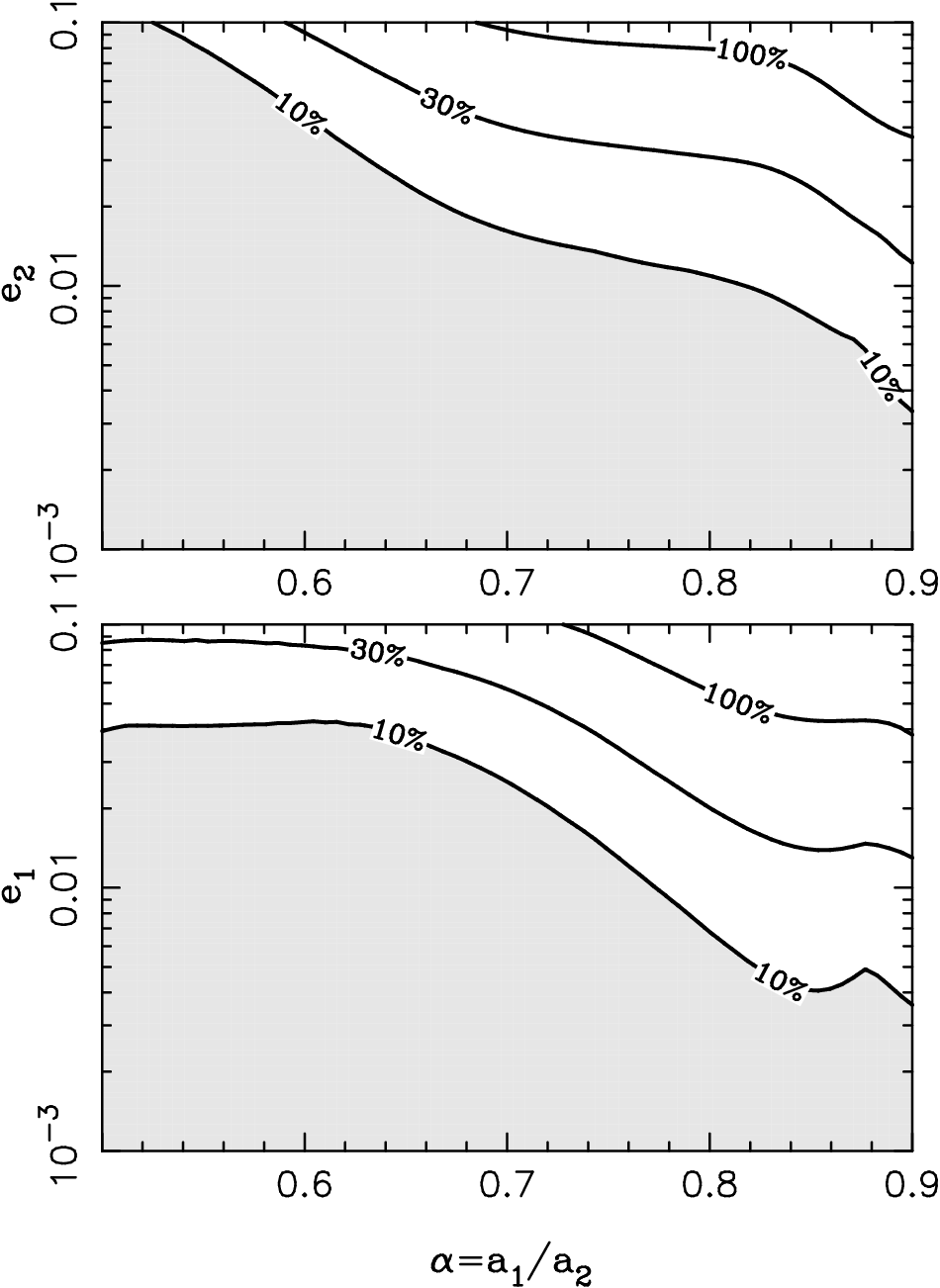}
\caption{The domain of validity of the analytical model as a function of $\alpha$,
$e_1$ and $e_2$. As in Figure \ref{comp1}, two planets with the mass $10^{-5}\, M_*$ 
each were placed on the initially coplanar orbits. The outer planet was set to 
have $a_2=1.0$ AU and the inner planet's semimajor axis was varied such that 
$0.5<\alpha<0.9$. The isolines show the relative precision of the analytic model.
The shaded area is where the precision of the analytic model is better than 10\%.}
\label{limit1}
\end{figure*}

\begin{figure*}
\epsscale{0.6}
\plotone{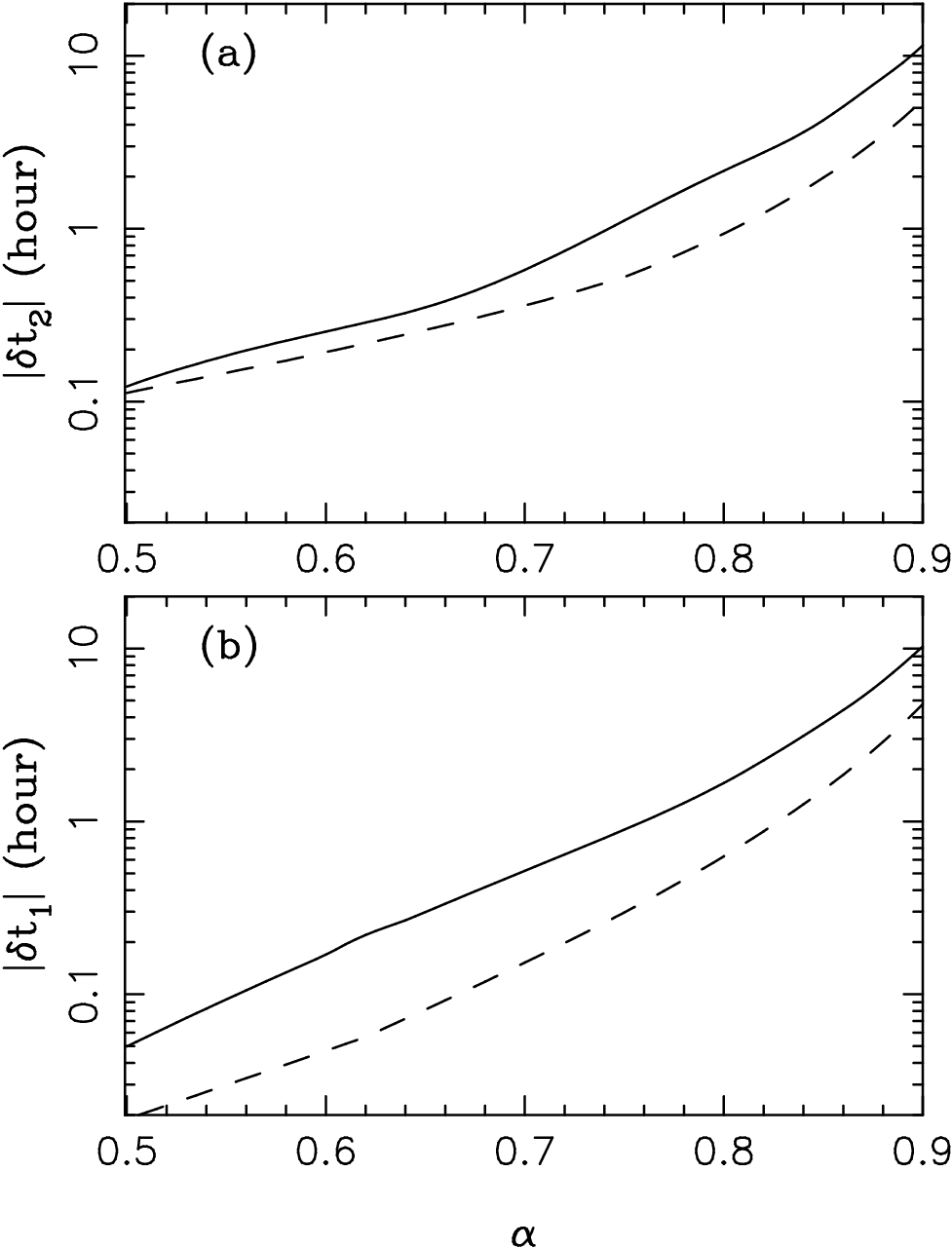}
\caption{The TTV amplitude of the outer (top  panel) and inner (bottom panel) planets.
The dashed lines show the amplitude of the step-like variation from $\delta \lambda_i/n_i$ (Eqs. 
\ref{dl1} and \ref{dl2}). The solid lines show the total amplitude from 
$(\delta \lambda_i+\delta \lambda_i^{\rm eff})/n_i$, where $\delta \lambda_i^{\rm eff}$ is given 
in Eqs. (\ref{dl1eff}) and (\ref{dl2eff}). Here we used $m_1=m_2=10^{-5}\ M_*$, $M_*=M_\odot$, $a_2=1$ AU, 
and $e_i^{(0)}=0$. The inner planet's semimajor axis was varied such that $0.5<\alpha<0.9$.
The TTV amplitudes are given here for a single conjunction between planets.}
\label{scale1}
\end{figure*}

\begin{figure*}
\epsscale{0.6}
\plotone{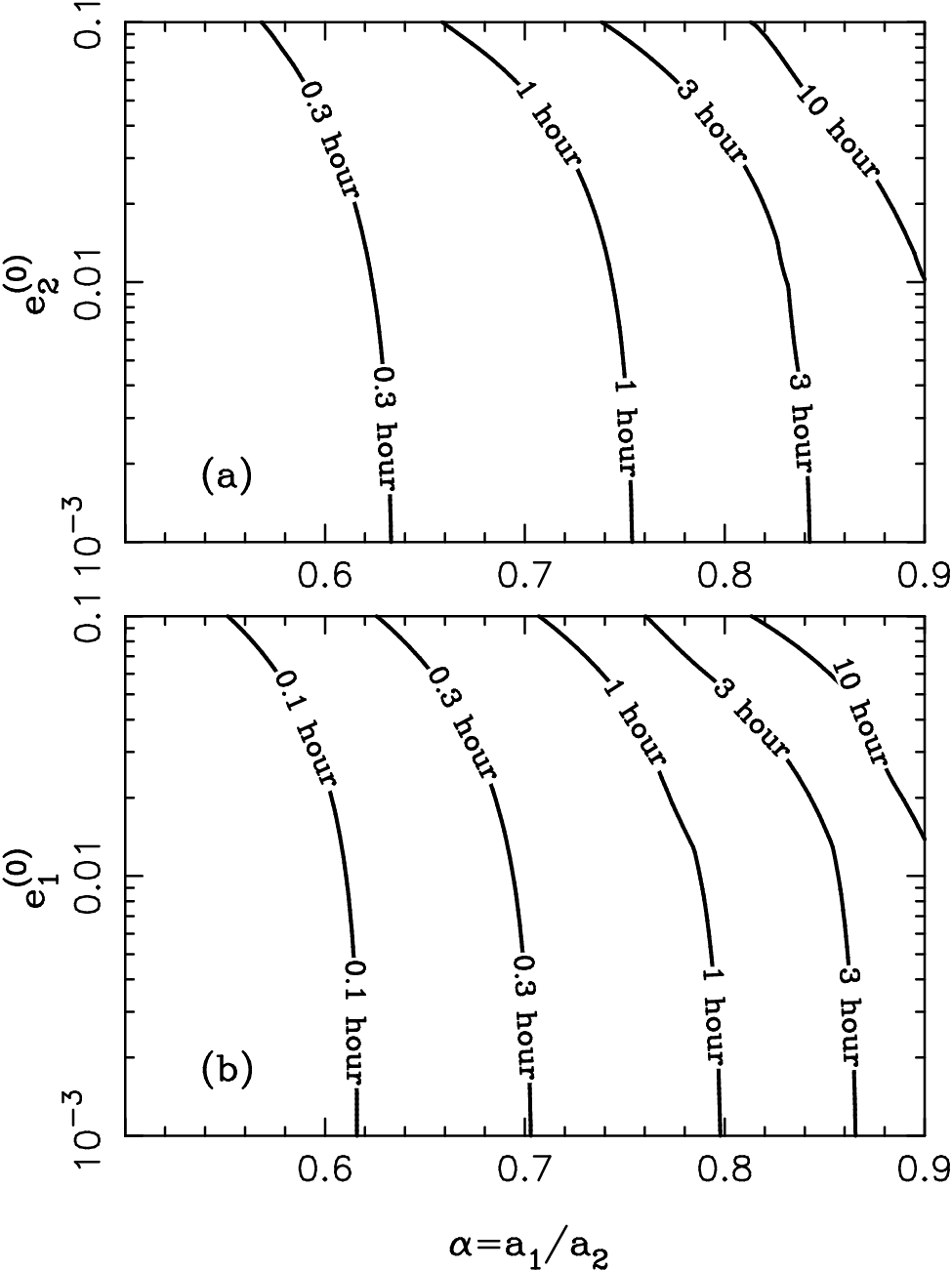}
\caption{The TTV amplitude of the outer (top  panel) and inner (bottom panel) planets.
We used $m_1=m_2=10^{-5}\ M_*$, $M_*=M_\odot$, $a_2=1$ AU, and varied $e_i^{(0)}$.
The inner planet's semimajor axis was chosen such that $0.5<\alpha<0.9$. The TTV amplitudes 
are given here for a single conjunction between planets.}
\label{scale2}
\end{figure*}

\begin{figure*}
\epsscale{0.6}
\plotone{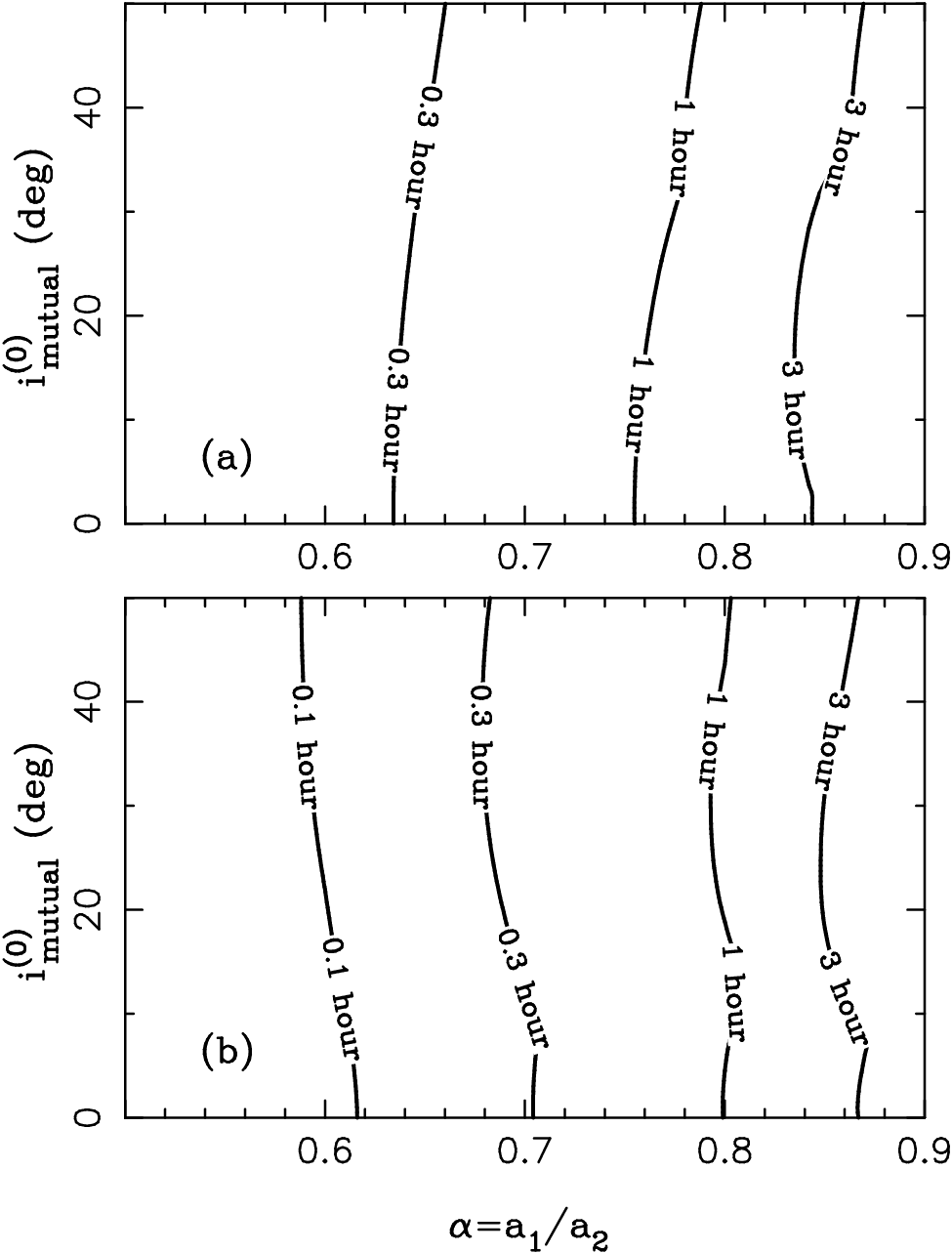}
\caption{The TTV amplitude of the outer (top  panel) and inner (bottom panel) planets.
We used $m_1=m_2=10^{-5}\ M_*$, $M_*=M_\odot$, $a_2=1$ AU, $e_i=0$ and varied the mutual
inclination between the two orbits. The inner planet's semimajor axis was chosen such that 
$0.5<\alpha<0.9$. The TTV amplitudes are given here for a single conjunction between 
planets.}
\label{scale3}
\end{figure*}

As a final word of caution, note that Eqs. (\ref{dt1}) or (\ref{dt2}) were derived under 
several assumptions. Most importantly, these equations are valid only for small orbital eccentricities. 
While many planetary systems will presumably fall into this category, perhaps the majority of them will not.
The TTVs for planetary systems with orbital eccentricities exceeding the threshold shown in Figure 
\ref{limit1} will contain many additional terms from the first and higher eccentricity powers. 
These terms will add frequencies $k_1 n_1 + k_2 n_2$, with arbitrary $k_1$ and $k_2$, potentially
generating resonant or near-resonant TTVs, and will modify the amplitude dependence of the $j(n_1 - n_2)$ 
frequencies on orbital parameters. The analytic model described here therefore cannot be  
used {\it in general} to characterize the planetary systems from TTVs. 

The main scientific value of the analytic model is to give us an intuitive framework for how the 
TTV method works in the limit of the nearly circular orbits. 

\acknowledgements
The work of DV was supported by Czech Grant Agency (grant P209-13-01308S). We thank the anonymous 
reviewer for very helpful comments.

\clearpage

\clearpage

\clearpage

\clearpage

\clearpage

\clearpage

\begin{thebibliography}

\bibitem{aetal05} Agol, E., Steffen, J., Sari, R., Clarkson, W., 2005,
 MNRAS, 359, 567

\bibitem{bc61} Brouwer, D., Clemence, G. M., 1961, Methods of celestial
 mechanics, Academic Press, 1961

\bibitem[Cabrera et al.(2014)]{2014ApJ...781...18C} Cabrera, J., Csizmadia, 
S., Lehmann, H., et al.\ 2014, \apj, 781, 18 

\bibitem[Carter et al.(2012)]{2012Sci...337..556C} Carter, J.~A., Agol, E., 
Chaplin, W.~J., et al.\ 2012, Science, 337, 556 

\bibitem[Deck et al.(2014)]{2014arXiv1403.1895D} Deck, K.~M., Agol, E., 
Holman, M.~J., \& Nesvorn\'y, D.\ 2014, arXiv:1403.1895 

\bibitem[Heyl 
\& Gladman(2007)]{2007MNRAS.377.1511H} Heyl, J.~S., \& Gladman, B.~J.\ 2007, \mnras, 377, 1511 

\bibitem[Holman 
\& Murray(2005)]{2005Sci...307.1288H} Holman, M.~J., \& Murray, N.~W.\ 2005, Science, 307, 1288 

\bibitem[Lithwick et al.(2012)]{2012ApJ...761..122L} Lithwick, Y., Xie, J., 
\& Wu, Y.\ 2012, \apj, 761, 122 

\bibitem{m93} Malhotra, R., 1993, ApJ, 407, 266

\bibitem{n09} Nesvorn\'y, D., 2009, ApJ, 701, 1116

\bibitem{nm08} Nesvorn\'y, D., Morbidelli, A., 2008, ApJ, 688, 636

\bibitem[Nesvorn{\'y} 
\& Beaug{\'e}(2010)]{2010ApJ...709L..44N} Nesvorn{\'y}, D., \& Beaug{\'e}, C.\ 2010, \apjl, 709, L44 

\bibitem[Nesvorn{\'y} et al.(2012)]{2012Sci...336.1133N} Nesvorn{\'y}, D., 
Kipping, D.~M., Buchhave, L.~A., et al.\ 2012, Science, 336, 1133 

\bibitem[Nesvorn{\'y} et al.(2013)]{2013ApJ...777....3N} Nesvorn{\'y}, D., 
Kipping, D., Terrell, D., et al.\ 2013, \apj, 777, 3 

\bibitem[Nesvorn{\'y} et al.(2014)]{2013Sci...336.1134N} Nesvorn{\'y}, D., 
Kipping, D., Terrell, D., et al.\ 2014, ApJ, in press

\bibitem[Nesvorn{\'y} et al.(2014)]{2012Sci...336.1134N} Vokrouhlick\'y, D.,  
Nesvorn\'y, D. 2014, submitted to ApJ

\end{thebibliography}
\end{document}